\documentclass[%
reprint, superscriptaddress,
 amsmath,amssymb,
 Aps, 
prd,
longbibliography,
]{revtex4-1}
\AtBeginDocument{%
    \newwrite\bibnotes
    \def\bibnotesext{Notes.bib}
    \immediate\openout\bibnotes=\jobname\bibnotesext
    \immediate\write\bibnotes{@CONTROL{REVTEX41Control}}
    \immediate\write\bibnotes{@CONTROL{%
    apsrev41Control,author="08",editor="1",pages="1",title="1",year="1"}}
     \if@filesw
     \immediate\write\@auxout{\string\citation{apsrev41Control}}%
    \fi
}%
\usepackage[utf8]{inputenc}
\usepackage{graphicx}
\usepackage{dcolumn}
\usepackage{bm}
\usepackage{xcolor, soul}
\sethlcolor{yellow}
\begin{document}

\preprint{APS/123-QED}

\title{Signatures of gate-driven out of equilibrium superconductivity\\
in Ta/InAs nanowires}

\author{Tosson Elalaily}
\affiliation{Department of Physics, Institute of Physics, Budapest University of Technology and Economics, M\"uegyetem rkp. 3., H-1111 Budapest, Hungary\\}
\affiliation{MTA-BME Superconducting Nanoelectronics Momentum Research Group, M\"uegyetem rkp. 3., H-1111 Budapest, Hungary\\}
\affiliation{Department of Physics , Faculty of Science, Tanta University, Al-Geish St., 31527 Tanta, Gharbia, Egypt\\}

\author{Martin Berke}
\affiliation{Department of Physics, Institute of Physics, Budapest University of Technology and Economics, M\"uegyetem rkp. 3., H-1111 Budapest, Hungary\\}
\affiliation{MTA-BME Superconducting Nanoelectronics Momentum Research Group, M\"uegyetem rkp. 3., H-1111 Budapest, Hungary\\}

\author{M\'at\'e Kedves}
\affiliation{Department of Physics, Institute of Physics, Budapest University of Technology and Economics, M\"uegyetem rkp. 3., H-1111 Budapest, Hungary\\}
\affiliation{MTA-BME Correlated van der Waals Structures Momentum Research Group, M\"uegyetem rkp. 3., H-1111 Budapest, Hungary\\}

\author{Gerg\H{o} F\"ul\"op}
\affiliation{Department of Physics, Institute of Physics, Budapest University of Technology and Economics, M\"uegyetem rkp. 3., H-1111 Budapest, Hungary\\}
\affiliation{MTA-BME Superconducting Nanoelectronics Momentum Research Group, M\"uegyetem rkp. 3., H-1111 Budapest, Hungary\\}

\author{Zolt\'an Scher\"ubl}
\affiliation{Department of Physics, Institute of Physics, Budapest University of Technology and Economics, M\"uegyetem rkp. 3., H-1111 Budapest, Hungary\\}
\affiliation{MTA-BME Superconducting Nanoelectronics Momentum Research Group, M\"uegyetem rkp. 3., H-1111 Budapest, Hungary\\}

\author{Thomas Kanne}
 \affiliation{Center for Quantum Devices and Nano-Science Center, Niels Bohr Institute, University of Copenhagen, Universitetsparken 5, DK-2100, Copenhagen, Denmark\\}
 
\author{Jesper Nygård}
 \affiliation{Center for Quantum Devices and Nano-Science Center, Niels Bohr Institute, University of Copenhagen, Universitetsparken 5, DK-2100, Copenhagen, Denmark\\}
 
\author{P\'eter Makk}
\email{makk.peter@ttk.bme.hu}
\affiliation{Department of Physics, Institute of Physics, Budapest University of Technology and Economics, M\"uegyetem rkp. 3., H-1111 Budapest, Hungary\\}
\affiliation{MTA-BME Correlated van der Waals Structures Momentum Research Group, M\"uegyetem rkp. 3., H-1111 Budapest, Hungary\\}

\author{Szabolcs Csonka}
\email{szabolcs.csonka@ttk.bme.hu}
\affiliation{Department of Physics, Institute of Physics, Budapest University of Technology and Economics, M\"uegyetem rkp. 3., H-1111 Budapest, Hungary\\}
\affiliation{MTA-BME Superconducting Nanoelectronics Momentum Research Group, M\"uegyetem rkp. 3., H-1111 Budapest, Hungary\\}

\begin{abstract}
Understanding the microscopic origin of the gate-controlled supercurrent (GCS) in superconducting nanobridges is crucial for engineering superconducting switches suitable for a variety of electronic applications. The origin of GCS is controversial, and various mechanisms have been proposed to explain it. In this work, we have investigated the GCS in a Ta layer deposited on the surface of InAs nanowires. Comparison between switching current distributions at opposite gate polarities and between the gate dependence of two opposite side gates with different nanowire$-$gate spacings shows that the GCS is determined by the power dissipated by the gate leakage. We also found a substantial difference between the influence of the gate and elevated bath temperature on the magnetic field dependence of the supercurrent. Detailed analysis of the switching dynamics at high gate voltages shows that the device is driven into the multiple phase slips regime by high-energy fluctuations arising from the leakage current.
\end{abstract}

\maketitle

\section*{Introduction}
Since superconducting circuits have the potential to realize electronics with short switching time and ultra-low power consumption, various architectures have been developed for integrating semiconductor technology with superconducting devices to reduce the high power consumption required for cooling the high-density semiconductor-based microchips \cite{mccaughan2014superconducting,mccaughan2019superconducting,frasca2019hybrid}.
The cryotron \cite{buck1956cryotron}, josephson cryotron \cite{matisoo1966subnanosecond}, rapid single flux quantum (RSFQ) device \cite{likharev1991rsfq}, and nanocryotron (nTron) \cite{mccaughan2014superconducting} were all developed as building blocks for superconducting switches, however, their scalability or even the difficulty of interfacing with CMOS electronics limited their applications.

In the recent years, suppression of supercurrent by applying a voltage to a gate electrode in the vicinity of superconducting metallic nanowire has attracted much attention as a promising building block for highly scalable superconducting switches \cite{de2018metallic,de2019josephson,paolucci2019field,paolucci2019magnetotransport,paolucci2019field2,de2020niobium,rocci2020gate,puglia2020electrostatic,puglia2020vanadium,bours2020unveiling,bours2020unveiling,puglia2021gate,de2021gate,puglia2021phase,orus2021critical,paolucci2021electrostatic,basset2021gate,elalaily2021gate,alegria2021high, ritter2021superconducting,golokolenov2021origin,catto2021microwave,ritter2022out}. In some works, the effect is attributed to the large electric field ($10^{8}$ V/m) at the superconducting surface \cite{de2018metallic,de2019josephson,paolucci2019field,paolucci2019magnetotransport,paolucci2019field2,de2020niobium,rocci2020gate,puglia2020electrostatic,puglia2020vanadium,bours2020unveiling,bours2020unveiling,puglia2021gate,de2021gate,puglia2021phase,orus2021critical,paolucci2021electrostatic}, which distorts the superconducting state and leads to the quenching of the superconductivity  \cite{mercaldo2020electrically,mercaldo2021spectroscopic,solinas2021sauter,chirolli2021impact, amoretti2022destroying}.
 Other studies \cite{alegria2021high,ritter2021superconducting,golokolenov2021origin,catto2021microwave,elalaily2021gate, basset2021gate,ritter2022out} reported a correlation between the gate controlled supercurrent (GCS) and the leakage current flowing between the gate and the superconducting device. Some of these studies suggest that the GCS results from ballistic injection of high-energy quasiparticles \cite{alegria2021high, ritter2021superconducting,golokolenov2021origin}. In another work, the quenching of the supercurrent was attributed to the absorption of phonons emitted in the relaxation process of high-energy electrons injected from the gate electrode \cite{ritter2022out}. In order to engineer efficient superconducting switches for future electronic applications, it is important to understand the dominant mechanism behind the CGS effect.

In this work, we have studied the GCS in a superconducting Ta shell deposited on the surface of InAs nanowires \cite{carrad2020shadow}. We chose Ta because of its strong spin-orbit interaction \cite{niimi2015reciprocal,velez2016hanle}, so it is expected that the electric field has a strong influence on the superconducting state. 
We investigated the influence of the distance between the gate and the nanowire on the suppression of the supercurrent for the fabricated devices. Also, the magnetic field dependence of the supercurrent under the influence of the gate voltage and elevated temperatures was investigated. In addition, the switching current distribution at opposite gate polarities and at different current ramp speeds was studied. Furthermore, we give a detailed analysis for the switching dynamics at high gate voltages. Our findings contradict the proposed theoretical explanations based on electric fields or ballistic injection of high-energy electrons, and they are consistent with the nonequilibrium phonons picture as the origin of the GCS effect.

\section*{Experiments}
\subsection{Device outline and characterization}
\begin{figure}[tb!]
	\includegraphics[width=\columnwidth]{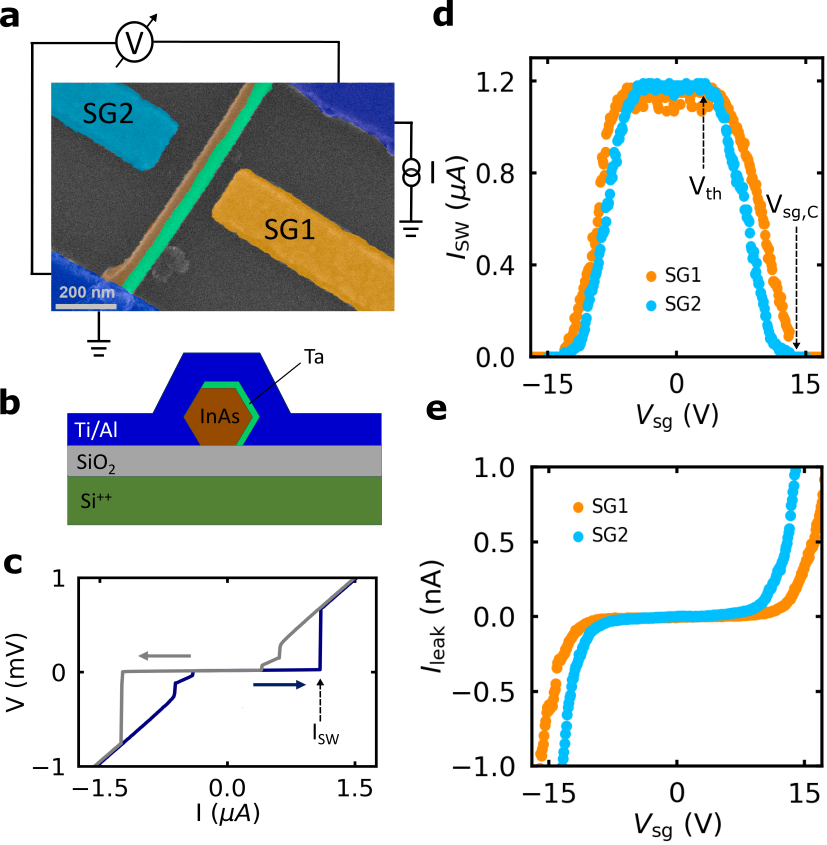}
	\caption{\label{fig:device}\textbf{Device geometry and gate dependence characterization (Device A)}. \textbf{a} A false-colored SEM image and \textbf{b} schematic of the side view of the nanowire device. \textbf{c} $I-V$ characteristics of the device measured at 35~mK. As the bias current ramps from negative to positive values (blue arrow), the device switches to finite-resistance state at the switching current $I_{\mathrm{SW}}$=~1.17~$\mu A$. If the current ramps in the opposite direction (grey arrow), the device switches back to the superconducting state at two successive retrapping current values at~$\simeq$~0.61~$\mu A$ and ~$\simeq$~0.4~$\mu A$. \textbf{d} $I_{\mathrm{SW}}$ as a function of $V_{\mathrm{sg}}$ (at magnetic field $B$~=~0.1~T) applied to SG1 (orange curve) and SG2 (light blue curve) with nanowire$-$gate spacings of~$\simeq$~65 and~$\simeq$~115 nm, respectively. \textbf{e} The leakage current as a function of $V_{\mathrm{sg}}$ for both gates.}
\end{figure}

\begin{figure*}[ht!]
	\includegraphics[width=\textwidth]{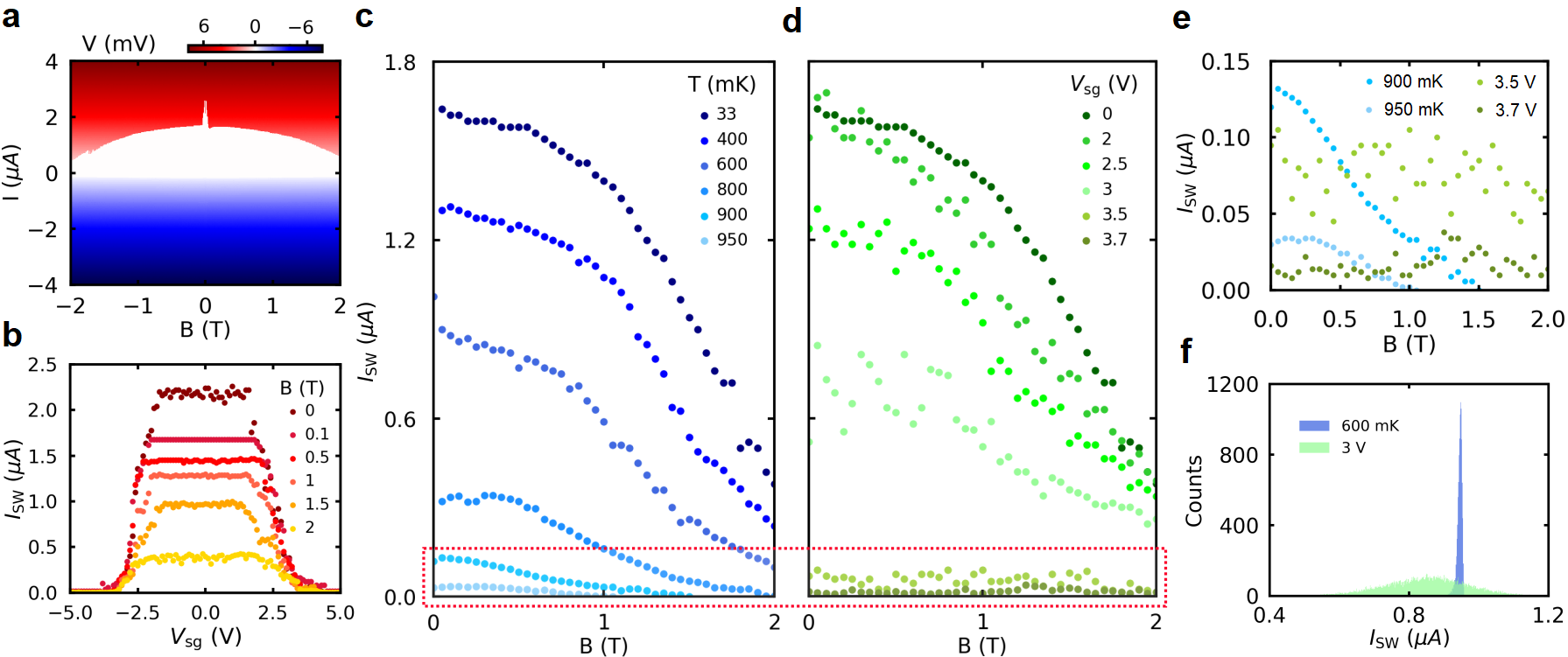}
	\caption{\label{fig:magnetdep}\textbf{Magnetic field dependence and comparison between the GCS effect and effect of bath temperature (device B, $d$ = 35 nm)}.  \textbf{a} $I-V$ curve as a function of out-of-plane magnetic field $B$ up to $\pm$~2~T. \textbf{b} $I_{\mathrm{SW}}$  as a function of $V_{\mathrm{sg}}$ at various values of B-field up to 2~T. \textbf{c} $I_{\mathrm{SW}}$ as a function of the B-field at various elevated temperatures and, \textbf{d} at various values of $V_{\mathrm{sg}}$. \textbf{e} Magnification of the curves surrounded by the red rectangle in \textbf{c} and \textbf{d}. \textbf{f} Comparison between the SCDs measured at T~=~600~mK (blue) and $V_{\mathrm{sg}}$~=~3 V (green).}  
\end{figure*}

\begin{figure*}[ht!]
	\includegraphics[width=\textwidth]{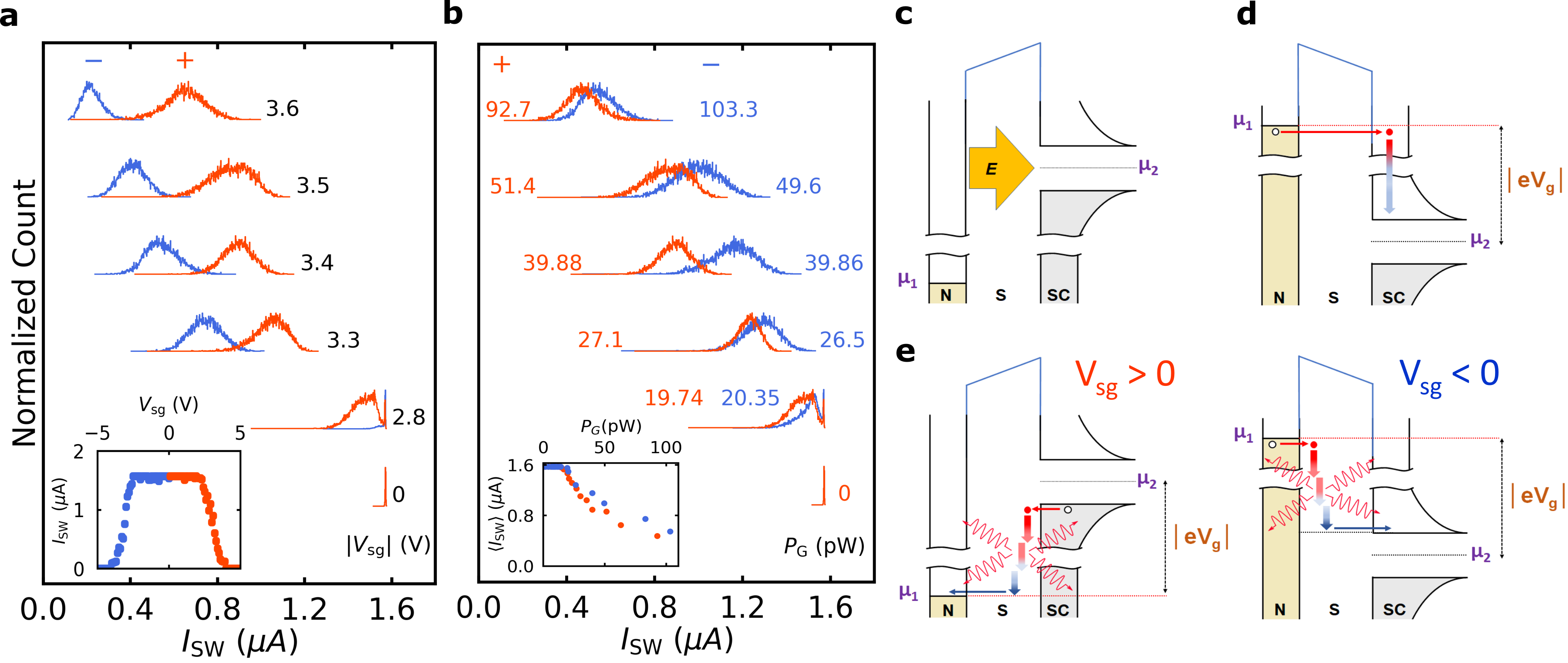}
	\caption{\label{fig:SCDpairing}\textbf{SCD measurements and schematics for different proposed mechanisms of the GCS effect (device C, $d$ = 30 nm)}. \textbf{a} SCDs measured at positive (orange) and negative (blue) gate polarity and paired at the same $|V_{\mathrm{sg}}|$. The SCDs are normalized to their maximum counts and shifted on the y-axis for clarity. The inset shows $I_{\mathrm{SW}}$ as a function of $V_{\mathrm{sg}}$ for the investigated device measured at 0.1 T. \textbf{b} SCDs measured at positive (orange) and negative (blue) gate polarity and paired at approximately the same $P_{\mathrm{G}}$. The inset shows the mean value of $\langle$$I_{\mathrm{SW}}$$\rangle$ of SCDs measured at both gate polarities as a function of $P_{\mathrm{G}}$. \textbf{c} Schematic diagram of the electric field $E$ applied from the metallic gate N to the superconducting nanowire SC at positive gate polarity. The colored/uncolored parts represent occupied/unoccupied states. \textbf{d} Schematic diagram of the ballistic electron injection from the gate to the nanowire at negative gate polarity. The high-energy electron (red circle) tunnels through the potential barrier of the substrate S and relaxes to the lowest unoccupied state (close to the superconducting gap edge), releasing heat on the SC-side. \textbf{e} Schematic diagram of relaxation of high-energy electrons in the substrate when injected from the SC/N side to N/SC at positive/negative gate polarity in the left/right panels. In the case of positive gate polarity, the electrons relax close to the SC side (superconducting nanowire) so that it is heated more than in the case of negative gate polarity at the same $P_{\mathrm{G}}$.}  
\end{figure*}

In our device configuration, we used InAs nanowires with a 20-nm-thick Ta shell layer deposited on only three facets of the nanowire \cite{carrad2020shadow}. In order to investigate the impact of the gate on the supercurrent flowing in the Ta layer, four-terminal nanowire-based devices were fabricated with the configuration shown in Fig.~\ref{fig:device}a, b. The Ta/InAs nanowires (green/brown) were deposited on a doped Si wafer with a 290-nm-thick oxide layer. Four Ti/Al contacts (blue) with a thickness of 10/80 nm were fabricated for quasi-four terminal measurements. A two metallic Ti/Au side gates SG1 (orange) and SG2 (light blue) with a thickness of 7/33~nm were placed with unequal spacings and on opposite sides of the nanowire. This provides a possibility to study the GCS effect for the device with gates at different spacings. The results presented in this paper are based on measurements performed on three different devices, A, B and C with the same device geometry, but with different values of nanowire$-$gate spacing $d$ in the range from 30 to 120 nm. The results in Fig.~\ref{fig:device} were measured on device A, in Fig.~\ref{fig:magnetdep} on device B, while the results in Fig.~\ref{fig:SCDpairing} and their analysis in Fig.~\ref{fig:escaperate} were performed on device C.

The current$-$voltage ($I-V$) characteristics measured at 35 mK show a clear switching from the superconducting state to the normal state at the switching current $I_{\mathrm{ SW }}$~$\simeq$~1.17~$\mu$A (see blue curve in Fig.~\ref{fig:device}c). When the measurements are carried out in the opposite sweep direction (grey curve), the device shows a hysteretic behaviour and switches back to the superconducting state at two successive retrapping current values at~$\simeq$~0.61~$\mu A$ and 0.4~$\mu A$. This hysteretic behaviour can be attributed to large Joule heating dissipated in the resistive state \cite{courtois2008origin}. The GCS is investigated by measuring the dependence of $I_{\mathrm { SW }}$ under the influence of gates SG1 (orange) and SG2 (light blue) with $d$ of~$\simeq$~65 and $\simeq$~115 nm, respectively. Fig.~\ref{fig:device}d shows $I_{\mathrm {SW}}$ as a function of $V_{\mathrm{sg}}$ for both gates, where each of the plotted curves has the same colour as the corresponding gate in Fig. 1a. The plot reveals that both gates completely switch the device to the normal state at almost the same critical gate voltage $V_{\mathrm{sg,C}}$ $\simeq$ $\pm$13 V. Despite the nanowire$-$gate spacing for SG1 is about half that for SG2, SG2 still suppresses $I_{\mathrm {SW}}$ at lower threshold gate voltage $V_{\mathrm{th}}$ than SG1. Importantly, at $V_{\mathrm{th}}$, a correspondingly large increase in the gate leakage current $I_{\mathrm{leak}}$ is observed for each of the gates (see Fig.~\ref{fig:device}e), which has also been reported elsewhere \cite{ritter2021superconducting,golokolenov2021origin,elalaily2021gate,basset2021gate,ritter2022out}.

\subsection{Magnetic field dependence}
The dependence of the supercurrent in our device on the out-of-plane magnetic field $B$ is shown in Fig.~\ref{fig:magnetdep}a, where the $I-V$ curves are measured as a function of the $B$-field up to $\pm$2~T. The white region represents the zero-resistance state, with a transition to and from the normal state (red and blue regions) at the switching and retrapping current values in the positive and negative bias current values, respectively. The magnitude of $I_{\mathrm{SW}}$ shows a rapid suppression with increasing $B$-field below 100~mT and then slowly decreases with further increasing the magnetic field up to 2~T. The sharp decrease in the critical current below 100~mT coincides with the $B_{\mathrm{C}}$ of the Al electrodes contacting the nanowire \cite{elalaily2021gate}, therefore we believe that this decrease is a result of the Al contacts switching to normal state. Although the maximum $B$-field in our setup (2~T) does not allow full suppression of the superconducting state in the Ta shell, based on the measured trend, $B_{\mathrm{C}}$ is expected to be about 3.5~T, which is consistent with earlier findings on identical Ta/InAs nanowires \cite{bjergfelt2019situ}.

The gate dependence of $I_{\mathrm{SW}}$ under the influence of B-field is shown in Fig.~\ref{fig:magnetdep}b. $I_{\mathrm{SW}}$ is plotted as a function of $V_{\mathrm{sg}}$ at different values of magnetic field up to 2~T. No significant change in $V_{\mathrm{sg,C}}$ with increasing $B$-field was observed, which is in contrast to the dependence observed for Ti and Al nanostructures \cite{de2018metallic,elalaily2021gate}. Fig.~\ref{fig:magnetdep}c and d show the dependence of $I_{\mathrm{SW}}$ on $B$-field under influence of temperature $T$ and $V_{\mathrm{sg}}$, respectively. In the former case, $I_{\mathrm{SW}}$ decreases with increasing $T$, as expected, accompanied by a suppression of $B_{\mathrm{C}}$, giving $B_{\mathrm{C}}$~=~2~T at 800~mK and $B_{\mathrm{C}}$~=~1.5~T at~900~mK. In the case of the gate control, $I_{\mathrm{ SW }}$ also decreases with increasing $V_{\mathrm{sg}}$, but surprisingly, $no$ $change$ $in$ $B_{\mathrm{C}}$ $was$ $observed$. For a better comparison, Figs.~\ref{fig:magnetdep}e shows a zoom in of the curves in both dependencies marked by the red rectangle and having almost the same magnitude of $I_{\mathrm{SW}}$ (at $B$~=~0~T). It can be clearly seen that the $B_{\mathrm{C}}$ dependence behaves differently under the influence of temperature and gate voltage. While from 900~mK to 950~mK $B_{\mathrm{C}}$ further decreases from 1.5~T to 1~T, $I_{\mathrm{SW}}$ does not seem to be suppressed by the magnetic field in the case of the gate (see also the Supporting Information) in strong contradiction to other works \cite{paolucci2019magnetotransport,bours2020unveiling,elalaily2021gate}.

Another remarkable difference between temperature and gate dependence is that $I_{\mathrm{ SW }}$ exhibits large fluctuations at finite gate voltages (see green curves in Fig.~\ref{fig:magnetdep}e). In order to investigate this effect, the switching current distribution (SCD) at finite temperatures and gate voltages is measured by ramping the current at constant speed from 0 to 3 $\mu$A for 10,000 times and recording the corresponding $I_{\mathrm{ SW }}$ value every time (see Methods). A comparison between the SCDs obtained at 600~mK and 3~V is shown in Fig.~\ref{fig:magnetdep}f. Despite the fact that both histograms have almost the same mean value $\langle$$ I_{\mathrm{SW}}$$\rangle$, the width of the histogram obtained under influence of the gate voltage is an order of magnitude larger than that obtained at elevated bath temperature. The large gate-induced broadening is consistent with Refs.~\citenum{puglia2020electrostatic,basset2021gate,golokolenov2021origin,puglia2021phase,ritter2022out} and shows that the gate voltage induces an out-of-equilibrium state in the superconducting nanowire, which cannot be described with an effective temperature.

 \subsection{Gate dependence of the probability distributions}
 
In the following, we will compare the SCDs measured at positive and negative gate polarity, as they are expected to behave differently for different microscopic origins of the GCS. The dependence of $I_{\mathrm{SW}}$ on $V_{\mathrm{sg}}$ of the device is shown in the inset of Fig.~\ref{fig:SCDpairing}a, where the positive and negative gate polarities represented by the orange and blue curves, respectively. Fig.~\ref{fig:SCDpairing}a shows the SCDs measured at the same $|V_{\mathrm{sg}}|$ but with opposite polarities are paired and shifted along the y-axis for clarity. For simplicity, we made the measurements with the Al leads in the normal state, at $B$~=~100~mT \cite{elalaily2021gate}. There is a clear difference in the shape and $\langle$$I_{\mathrm{ SW }}$$\rangle$ of SCDs paired at equal $|V_{\mathrm{sg}}|$. In addition, we also paired SCDs for opposite gate polarities and with approximately the same power dissipated at the gate $P_{\mathrm{G}}$~=~$I_{\mathrm{leak}}$$\cdot$$V_{\mathrm{sg}}$ as shown in Fig.~\ref{fig:SCDpairing}b. Comparing Fig.~\ref{fig:SCDpairing}a and b, one can conclude that the pairing at the same power gives a better match between SCDs with opposite polarities. We also found that the SCDs measured at positive polarity have a slightly smaller $\langle$$ I_{\mathrm{ SW }}$$\rangle$ than those measured at negative polarity at the same $P_{\mathrm{G}}$ (see inset in Fig.~\ref{fig:SCDpairing}b).

Assuming that the electric field $E$ applied by the gate (Fig.~\ref{fig:SCDpairing}c) is responsible for the suppression of $I_{\mathrm{ SW }}$ \cite{mercaldo2020electrically,mercaldo2021spectroscopic,solinas2021sauter,chirolli2021impact, amoretti2022destroying}, its effect should not depend on the sign of $E$. Therefore, we expect the SCD obtained at a given voltage $V_{\mathrm{sg}}$ to be identical to the SCD obtained at the same gate voltage with the opposite sign, -$V_{\mathrm{sg}}$. Since the measured SCDs do not match at opposite polarities (see Fig.~\ref{fig:SCDpairing}a), our results contradict the electric field-based explanation. Another possible microscopic picture is that the CGS is caused by ballistic injection of high-energy quasiparticles, as shown in Fig.~\ref{fig:SCDpairing}d. After injection of these electrons, their energy is released by relaxation, heating the side on which they end up. Therefore, for negative gate polarity (Fig.~\ref{fig:SCDpairing}d), they heat the superconducting bridge, while for positive polarity they heat the gate electrode instead.
Thus, a stronger suppression of superconductivity is expected for negative polarity. Therefore, at the same $P_{\mathrm{G}}$ value, the mean value of the distribution is expected to be significantly smaller for negative polarity than for positive polarity. Comparing this prediction with the measured results in Fig.~\ref{fig:SCDpairing}b, one can conclude that the experimental findings are just opposite, so that ballistic injection of electrons can also be excluded. 

The most likely explanation for our results is the generation of phonons by series of relaxation events of the high-energy electrons in the substrate \cite{ritter2022out}. 
The small shift between the $\langle$$I_{\mathrm{SW}}$$\rangle$ measured for the two polarities (see Fig.~\ref{fig:SCDpairing}b) can be attributed to the short energy relaxation length of electrons in SiO$_{\mathrm{2}}$ ($\leq3$ nm) at high electric fields compared to nanowire$-$gate spacing ($d=30$ nm) \cite{fischetti1985theory,brorson1985direct,fischetti1987ballistic,dimaria1988vacuum}. Thus, at positive gate polarity, it is expected that the high-energy electrons will relax close to the nanowire (Fig.~\ref{fig:SCDpairing}e, left panel) and the generated phonons can heat the superconducting nanowire more than at negative gate polarity (Fig.~\ref{fig:SCDpairing}e, right panel).

 \subsection{Analysis of the switching dynamics}

\begin{figure}[ht!]
	\includegraphics[width=\columnwidth]{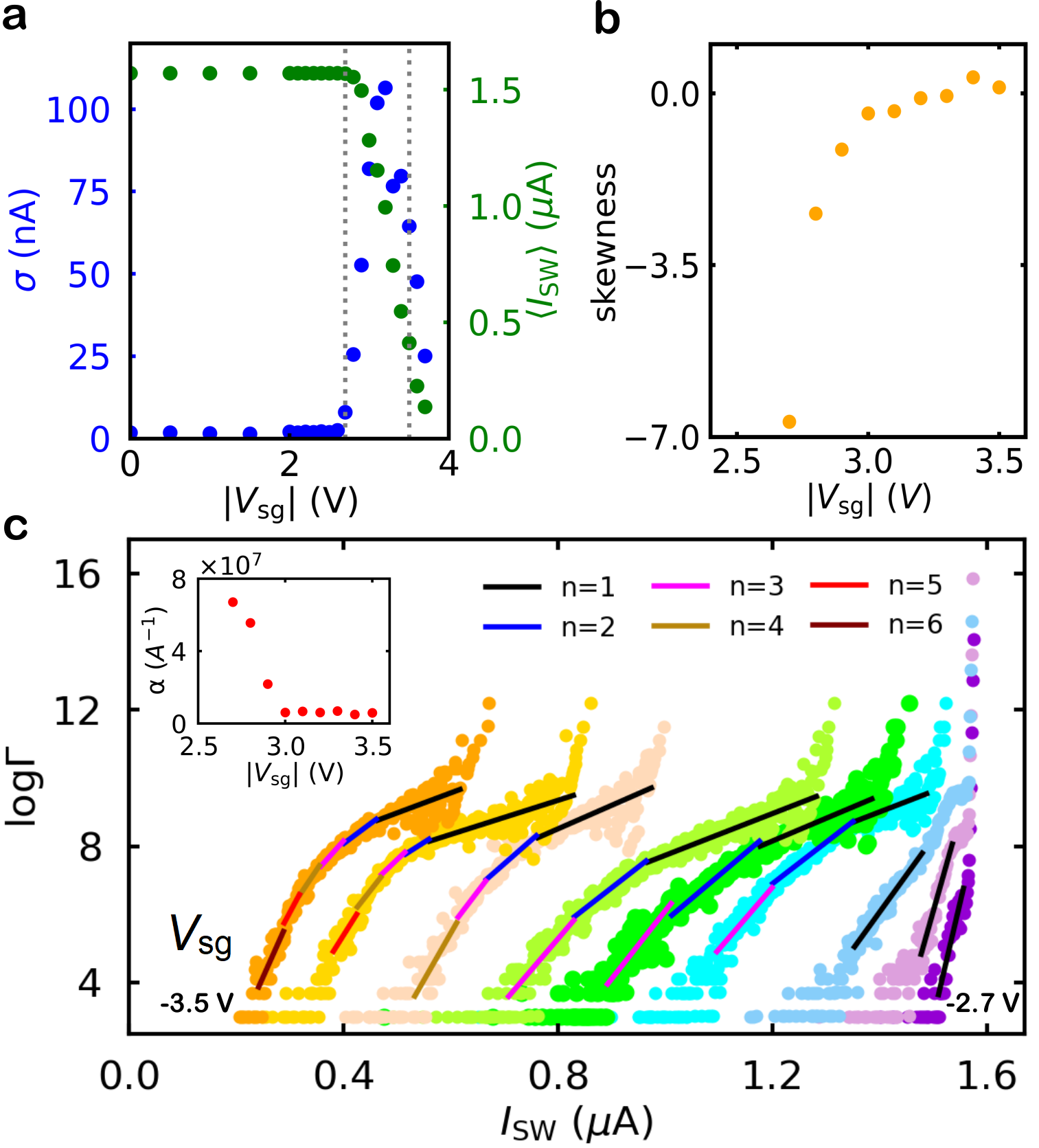}
	\caption{\label{fig:escaperate}\textbf{Analysis of the switching dynamics under influence of $V_{\mathrm{sg}}$ (device C, $d$ = 30 nm)}. \textbf{a} Standard deviation $\sigma$ and mean value $\langle$$I_{\mathrm{SW}}$$\rangle$ as a function of $|V_{\mathrm{sg}}|$ for all SCDs measured at negative gate polarity in the blue and green curves, respectively. \textbf{b} The calculated skewness as a function of $|V_{\mathrm{sg}}|$ for SCDs measured with a step of 0.1 V in the interval [-3.5, -2.7 V] (surrounded by the vertical grey dotted lines in panel a) where a corresponding increase in $I_{\mathrm{leak}}$ is observed. \textbf{c} Logarithm of escape rate $\Gamma$ as a function of $I_{\mathrm{ SW }}$ (colored curves) for different values of $V_{\mathrm{sg}}$ from -3.5 V (orange curve) up to -2.7 V (purple curve) with a step of 0.1 V. The colored solid lines represent the fitting of different portions of these curves with an exponential of higher orders n of the slope $\alpha$. The inset shows the variation of the slope $\alpha$ with increasing $V_{\mathrm{sg}}$.}  
\end{figure}

The standard deviation $\sigma$ of SCDs measured under the influence of the gate is represented by the blue curve in Fig.~\ref{fig:escaperate}a. For small values of $|V_{\mathrm{sg}}|$, where $I_{\mathrm{leak}}$ is negligible, $\sigma$ is independent of $|V_{\mathrm{sg}}|$ and no significant change in the $\langle$$ I_{\mathrm{SW}}$$\rangle$ of SCDs (green curve) was observed. Beyond $V_{\mathrm{th}}$ at $|V_{\mathrm{sg}}|$~=~2.7~V, $\sigma$~ increases with $|V_{\mathrm{sg}}|$ because the fluctuations assisted by $I_{\mathrm{leak}}$ become stronger and more frequent. This increases the probability of nanowire switching at small $I_{\mathrm{ SW }}$ values with a corresponding suppression in the $\langle$$I_{\mathrm{ SW }}$$\rangle$ of the SCD. This increase in the width of the SCDs is analogous to the typical temperature dependence (see the Supporting Information) \cite{ejrnaes2019superconductor,puglia2020electrostatic,puglia2021phase} associated with thermally-activated phase slips \cite{mccumber1970time,bezryadin2013superconductivity}. However, the large width of the SCDs obtained under the influence of the gate indicates that the system is driven to a nonequilibrium state where the fluctuations are an order of magnitude larger than expected from the bath temperature. With further increasing $|V_{\mathrm{sg}}|$, $\sigma$ decreases and the SCDs become more symmetric, as shown by their calculated skewness in Fig.~\ref{fig:escaperate}b. This is analogous with the picture that the switching of the system is due to multiple phase slips (MPS) found at finite temperatures \cite{ejrnaes2019superconductor}.

Interestingly, the SCD in Fig.~\ref{fig:SCDpairing}a  at $V_{\mathrm{sg}}$~=~2.8 V (orange curve) shows two peaks, a sharp one at 1.57 $\mu$A and a broad one around 1.5 $\mu$A. This distribution looks like the sum of two overlapping probability distributions, similar to distributions shown in Refs.~\citenum{puglia2020electrostatic,basset2021gate}. Since the probability distribution in this transition region depends strongly on the ramp speed of the bias current $\nu_{\mathrm{I}}$ (see the Supporting Information), we could completely switch between the two distributions when the ramp speed was changed from 300 (at which the SCDs in Fig.~\ref{fig:SCDpairing} are measured) to 9.375 $\mu$A/s. For a more accurate evaluation, it is better to transform the measured probability distributions into the speed-independent escape rate $\Gamma(I,T)$ (see the Supporting Information) by the direct Kurkijärvi-Fulton-Dunkleberger (KFD) transformation \cite{kurkijarvi1972intrinsic,fulton1974lifetime,bezryadin2013superconductivity}:
 
\begin{equation}
\label{escape rate}
\Gamma(I_{\mathrm{N}},T)=\frac{P(I_{\mathrm{N}},T)\nu_{\mathrm{I}}}{1-w\sum_{k=0}^{N}P(I_{\mathrm{k}},T)}
\end{equation} 
where $w$ is the bin size in the current axis of the measured probability distribution $P(I,T)$, and $P(I_{\mathrm{k}},T)$ is the switching probability in the bias current interval $[kw,(k + 1)w]$ with $k$ $\in$ $[0,N]$. Fig.~\ref{fig:escaperate}c shows the logarithm of the calculated $\Gamma(I_{\mathrm{N}},T)$ as a function of current for SCDs measured under the influence of $V_{\mathrm{sg}}$ in the interval [-3.5, -2.7 V]. As long as $V_{\mathrm{sg}}$ is small, the SCDs have a sharp peak around $I_{\mathrm{ SW }}$=1.57 $\mu$A, resulting in a large escape rate around this value, which represents the escape rate due to quantum tunnelling or thermal escape~$\Gamma_{\mathrm{T}}$. As the influence of the gate voltage sets in (see e.g., purple curve), a finite escape rate appears at lower $I_{\mathrm{ SW }}$ values, corresponding to the gate-assisted escape rate~$\Gamma_{\mathrm{L}}$. The latter contribution becomes the dominant escape rate at higher gate voltages (see e.g., green curve).

For $V_{\mathrm{sg}}$ $>$~-3 V, where $I_{\mathrm{leak}}$ is negligible (the first three curves from the right), $\Gamma_{\mathrm{L}}$ can be well fitted with an exponential curve (black solid line) given by

\begin{equation}
\label{fitting exponential}
\Gamma_{L}(I_{\mathrm{sw}})=Ae^{n\alpha I_{\mathrm{sw}}}
\end{equation} 
with $n=1$ and using $\alpha$ and $A$ as fitting parameters. The switching dynamics in this region have been studied in detail in Ref.~\citenum{basset2021gate}. On the other hand, for $V_{\mathrm{sg}}$~$\leq$~-3~V $\Gamma_{\mathrm{L}}$ deviates from the single exponential dependence described by Eq.~\ref{fitting exponential}. For example, the light green curve measured at -3.2 V can be fitted at large current values using Eq.~\ref{fitting exponential} with n=1 (see black solid lines). Interestingly, the measured curve for $I_{\mathrm{SW}}$ $<$ 0.9 $\mu$A can be well fitted by adding extra higher order terms with n=2,3,$\ldots$, keeping the same values of the fitting parameters. Further increasing $V_{\mathrm{sg}}$ ($I_{\mathrm{leak}}$) requires more higher order terms to fit the escape rate dependence (e.g. orange curve) \cite{ejrnaes2019superconductor}. The value of $\alpha$ required to fit the escape rate dependence decreases sharply as $V_{\mathrm{sg}}$ increases, and saturates at large values of $V_{\mathrm{sg}}$ as shown in the inset of Fig.~\ref{fig:SCDpairing}c.

The deviation of the escape rate dependence with current from a pure exponential at higher gate voltages is similar to elevated temperatures in Ref.~\citenum{ejrnaes2019superconductor}. This can be attributed to the reduced impact for a single fluctuation event triggered by the leakage current, since the dissipation during the induced phase slip event is smaller at lower values of $I_{\mathrm{SW}}$ \cite{bezryadin2013superconductivity,ejrnaes2019superconductor}. Thus, several coincident fluctuation events with corresponding induced multiple phase slips (MPS) are required to trigger the switching of the nanowire to the normal state \cite{ejrnaes2019superconductor,shah2008inherent,pekker2009stochastic}. In this regime, at large bias current values, the dissipation of a single MPS event ($n=1$) is sufficient to switch the nanowire into the resistive state. On the other hand, at lower current values, the dissipation of a single MPS event is reduced and higher orders ($n=2,3,...$) of the MPS event are required to trigger the resistive switching of the nanowire \cite{ejrnaes2019superconductor}.

\section*{Comparison with the proposed microscopic pictures}
In the following, we will compare our experimental results with the possible microscopic pictures. Starting from the two gates (Fig.~\ref{fig:device}d,e), despite SG2 has almost twice the nanowire$-$gate spacing of SG1, it suppresses the $I_{\mathrm{ SW }}$ at lower $V_{\mathrm{th}}$ than SG1. This contradicts the electric field picture as a possible explanation for the origin of the GCS. On the other hand, $I_{\mathrm{ SW }}$ starts to be suppressed with the onset of leakage current between the nanowire and each of the gates (see the Supporting Information). In another cool-down, the influence of the two gates for the same device shows an opposite situation, as SG1 shows a stronger influence on $I_{\mathrm{SW}}$ than SG2 (see the supporting information). This excludes any concerns arising from the quite large dielectric constant of the InAs nanowire between SG2 and the Ta shell (see Fig.~\ref{fig:device}a), which may lead to a larger influence of SG2 on $I_{\mathrm{SW}}$ than SG1. Interestingly, we found that the influence of the two gates on $I_{\mathrm{ SW }}$ gives better matching with $P_{\mathrm{G}}$ in the two cool-downs (see the Supporting Information).

Accepting that the leakage current plays a key role in the GCS, a simple explanation arises: that the leaking electrons increase the temperature of the superconducting nanowire. We have investigated for the first time the $B$-field dependence of superconducting nanowire with normal contacts, which allows efficient cooling of the superconducting nanowire. The $B$-field dependence at finite $T$ and finite gate voltage was strictly different, indicating that the effect of leakage current cannot be described by simple hot electron regime induced by elevated bath temperature. The highly non-equilibrium state of the superconductor at finite gate voltage is further supported by the broad SCDs in our work and in previous results \cite{puglia2020electrostatic,basset2021gate,golokolenov2021origin,puglia2021phase,ritter2022out}. Our detailed comparison of the SCDs for different gate polarities (Fig.~\ref{fig:SCDpairing}) provided another important finding which is inconsistent with electric-field-induced suppression of superconductivity. Pairing of SCDs measured at opposite gate polarities at the same leakage current dissipation, $P_{\mathrm{G}}$, provided a better matching than at the same $|V_{\mathrm{sg}}|$ (Fig.~\ref{fig:SCDpairing}a,b). This reveals that the suppression of $I_{\mathrm{SW}}$ depends not only on the energy of the injected electrons (e$V_{\mathrm{sg}}$) or the rate of their injection ($I_{\mathrm{leak}}$/e), but on the power dissipated at the gate $P_{\mathrm{G}}$. Based on the $\langle$$I_{\mathrm{SW}}$$\rangle$ of the SCDs for the two polarities, the ballistic injection of electrons from the gate into the superconducting nanowire can be discarded. We conclude that the phonon-mediated excitation of the superconductor remains a microscopic picture consistent with the measured results.

Furthermore, we also noticed that the power dissipation at the gate required to fully suppress $I_{\mathrm{ SW }}$, $P_{\mathrm{G,C}}$ is comparable to the power dissipation that occurs when the device switches to the resistive state $P_{\mathrm{n}}$~=~$I_{\mathrm{ SW }}^{\mathrm{2}}$$\cdot$$R_{\mathrm{n}}$. For example, for device A, in the case of SG1 (the closer to the Ta shell), $P_{\mathrm{G,C}}$~$\simeq$~1.5~nW (see the Supporting Information), while $P_{\mathrm{n}}$~$\simeq$~ 1 nW (using $R_{\mathrm{n}}$= 780 $\Omega$ and $I_{\mathrm{ SW }}$~$\simeq$~1.17~$\mu$A).

 Finally, a very large leakage current was required to quench $I_{\mathrm{SW}}$ when we investigated the GCS in similar Ta/InAs devices fabricated on a sapphire substrate (see the Supporting Information). These results indicate that the GCS depends mainly on the properties of the substrate and the leakage pathway between the gate and nanowire.

\section*{Conclusions}
We investigated the origin of GCS in the Ta half-shell layer deposited on InAs nanowires by various measurements. Devices with small nanowire-gate spacing (specifically devices B and C) fully switch to the normal state below $V_{\mathrm{sg}}$~=~$\pm$5~V, which makes them promising for integration into classical electronic circuits. When the wire is connected by electrodes in the normal state, the critical magnetic field $B_{\mathrm{C}}$ is not suppressed under the influence of the gate as for elevated temperatures. Moreover, the comparison of the switching current distributions at opposite gate polarities, as well as the gate dependence of two opposite side gates at different nanowire-gate spacings show that the power dissipated at the gate $P_{\mathrm{G}}$ is the relevant parameter for this effect. Analysis of the switching dynamics under strong gate influence shows a deviation in the escape rate dependence with the bias current from a pure exponential. This indicates that the device is driven into the MPS regime by the high energy fluctuations originating from the leakage current. Our findings contradict the microscopic pictures proposing electric fields or ballistic injection of high-energy electrons as the origin of the GCS effect, but they are consistent with the non-equilibrium superconducting state resulting from the absorption of phonons generated by the leakage current.

\section*{Methods}
InAs nanowires were grown by the VLS mechanism using Molecular Beam Epitaxy and the Ta shell was deposited in-situ under UHV using electron beam evaporation at a substrate temperature of about $25^{\circ}$~C. Based on the TEM characterization, the morphology of the Ta shells was continuous but granular on the InAs nanowires and was found to be non-crystalline \cite{bjergfelt2019situ}.

The Ta/InAs nanowires were deposited on the top of a doped Si wafer with 290~nm thick $\mathrm{SiO_{2}}$ layer by means of a hydraulic micromanipulator along with a high magnification optical microscope. The nanowire device was fabricated in two separate electron beam lithography (EBL) steps. In the first step, four Ti/Al contacts with a thickness of 10/80~nm were fabricated. Prior to the metal evaporation, Ta/InAs nanowires were exposed to Ar-ion plasma milling for 8 minutes at 50 W to remove any oxides on the top of the Ta shell. In the second step, two metallic gates of Ti/Au layers with a thickness of 7/33~nm were fabricated with unequal spacing and on opposite sides of the nanowire. The metallic gates were fabricated in a separate lithography step, since thin resist is used for precise alignment of the gates from the nanowire.

The $I-V$ characteristics of the device were measured by a pure DC measurements using a quasi 4-probe method in which the current was injected through the nanowire via a pair of Al contacts by using a standard voltage source (Basel DAC SP 927) with series resistor of 1~M$\Omega$, while the voltage was measured across the other pair with a differential voltage amplifierand a digital multimeter (Keithley 2001). The leakage current was recorded by measuring the voltage across a 10~M$\Omega$ preresistor connected to the gate and corrected according to the method reported in Ref.~\citenum{elalaily2021gate}. 

The SCD was measured using a NI-DAQ card (USB-6341), where a periodic current wave signal was engineered. This signal is composed of a positive linear ramp with an amplitude of 3~$\mu$A and a slope in the range from 9.375 to 300~$\mu$A/s followed by a 2.5 ms of zero-current plateau for cooling down the superconducting device. This signal is repeated 10,000 times, and $I_{\mathrm{SW}}$ is extracted each time. All SCDs are measured at 0.1~T to switch the Al leads to the normal state. All measurements were carried out in a Leiden Cryogenics CF-400 top-loading cryo-free dilution refrigerator system with a base temperature of 30~mK.

\section*{Author contributions}
 T.E., M.B. fabricated the devices, T.E., M.B., M.K., G.F. and Z.S. performed the measurements and did the data analysis. T. K. and J. N. developed the nanowires. All authors discussed the results and worked on the manuscript. P.M. and S.C. guided the project.

\section*{Acknowledgments}
This work has received funding from Topograph FlagERA, the SuperTop QuantERA network, SuperGate Fet Open, the FET Open AndQC, and from the OTKA FK-123894 grants.
This research was supported by the Ministry of Innovation and Technology and the National Research, Development and Innovation Office within the Quantum Information National Laboratory of Hungary and by the Quantum Technology National Excellence Program (Project Nr. 2017-1.2.1-NKP-2017-00001), by the \'UNKP-22-5 New National Excellence Program, the János Bolyai Research Scholarship of the Hungarian Academy of Sciences, by ÚNKP-22-2-I-BME-22 New National Excellence Program of the Ministry for Culture and Innovation from the source of the National Research, Development and Innovation Fund, by the Carlsberg
Foundation, Innovation Fund Denmark and the Danish National Research Foundation. We thank M. Bjergfelt, D. Carrad and T.S. Jespersen for their contributions to the development of the hybrid nanowires.

\bibliography{references}

\end{document}


\title{Signatures of gate-driven out of equilibrium superconductivity\\
in Ta/InAs nanowires}

\author{Tosson Elalaily}
\affiliation{Department of Physics, Institute of Physics, Budapest University of Technology and Economics, M\"uegyetem rkp. 3., H-1111 Budapest, Hungary\\}
\affiliation{MTA-BME Superconducting Nanoelectronics Momentum Research Group, M\"uegyetem rkp. 3., H-1111 Budapest, Hungary\\}
\affiliation{Department of Physics , Faculty of Science, Tanta University, Al-Geish St., 31527 Tanta, Gharbia, Egypt\\}

\author{Martin Berke}
\affiliation{Department of Physics, Institute of Physics, Budapest University of Technology and Economics, M\"uegyetem rkp. 3., H-1111 Budapest, Hungary\\}
\affiliation{MTA-BME Superconducting Nanoelectronics Momentum Research Group, M\"uegyetem rkp. 3., H-1111 Budapest, Hungary\\}

\author{M\'at\'e Kedves}
\affiliation{Department of Physics, Institute of Physics, Budapest University of Technology and Economics, M\"uegyetem rkp. 3., H-1111 Budapest, Hungary\\}
\affiliation{MTA-BME Correlated van der Waals Structures Momentum Research Group, M\"uegyetem rkp. 3., H-1111 Budapest, Hungary\\}

\author{Gerg\H{o} F\"ul\"op}
\affiliation{Department of Physics, Institute of Physics, Budapest University of Technology and Economics, M\"uegyetem rkp. 3., H-1111 Budapest, Hungary\\}
\affiliation{MTA-BME Superconducting Nanoelectronics Momentum Research Group, M\"uegyetem rkp. 3., H-1111 Budapest, Hungary\\}

\author{Zolt\'an Scher\"ubl}
\affiliation{Department of Physics, Institute of Physics, Budapest University of Technology and Economics, M\"uegyetem rkp. 3., H-1111 Budapest, Hungary\\}
\affiliation{MTA-BME Superconducting Nanoelectronics Momentum Research Group, M\"uegyetem rkp. 3., H-1111 Budapest, Hungary\\}

\author{Thomas Kanne}
 \affiliation{Center for Quantum Devices and Nano-Science Center, Niels Bohr Institute, University of Copenhagen, Universitetsparken 5, DK-2100, Copenhagen, Denmark\\}
 
\author{Jesper Nygård}
 \affiliation{Center for Quantum Devices and Nano-Science Center, Niels Bohr Institute, University of Copenhagen, Universitetsparken 5, DK-2100, Copenhagen, Denmark\\}
 
\author{P\'eter Makk}
\email{makk.peter@ttk.bme.hu}
\affiliation{Department of Physics, Institute of Physics, Budapest University of Technology and Economics, M\"uegyetem rkp. 3., H-1111 Budapest, Hungary\\}
\affiliation{MTA-BME Correlated van der Waals Structures Momentum Research Group, M\"uegyetem rkp. 3., H-1111 Budapest, Hungary\\}

\author{Szabolcs Csonka}
\email{szabolcs.csonka@ttk.bme.hu}
\affiliation{Department of Physics, Institute of Physics, Budapest University of Technology and Economics, M\"uegyetem rkp. 3., H-1111 Budapest, Hungary\\}
\affiliation{MTA-BME Superconducting Nanoelectronics Momentum Research Group, M\"uegyetem rkp. 3., H-1111 Budapest, Hungary\\}

\maketitle

 \section{gate dependence of the supercurrent for devices B and C} 
\begin{figure}[ht!]
	\includegraphics[width=0.8\columnwidth]{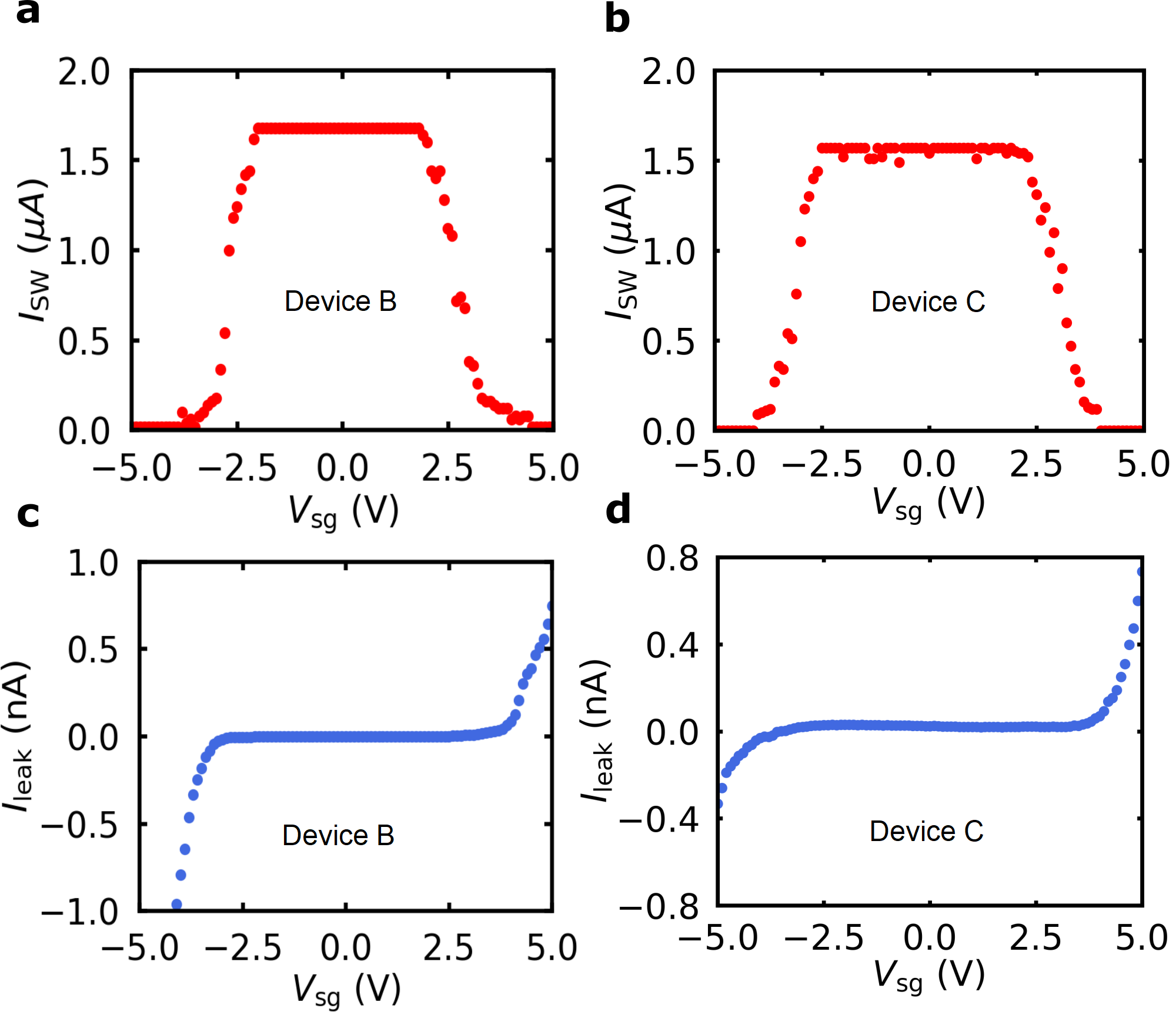}
	\caption{\label{fig:depend_gate}\textbf{a},\textbf{b} $I_{\mathrm{SW}}$ as a function of $V_{\mathrm{sg}}$ for devices B and C  at $B~=0.1$~T and their corresponding leakage current in \textbf{c},\textbf{d}, respectively.}
\end{figure}

SFig.~\ref{fig:depend_gate}a and b show the $I_{\mathrm{SW}}$ as a function of $V_{\mathrm{sg}}$ for devices B and C, respectively. The gate dependence of the devices was investigated by using SG1 (the closest to Ta shell) at $B=0.1$~T. Panels c and d show the corresponding $I_{\mathrm{leak}}$ as a function of $V_{\mathrm{sg}}$. 
  
\section{Magnetic field dependence under influence of the gate} 
\begin{figure}[ht!]
	\includegraphics[width=0.6\columnwidth]{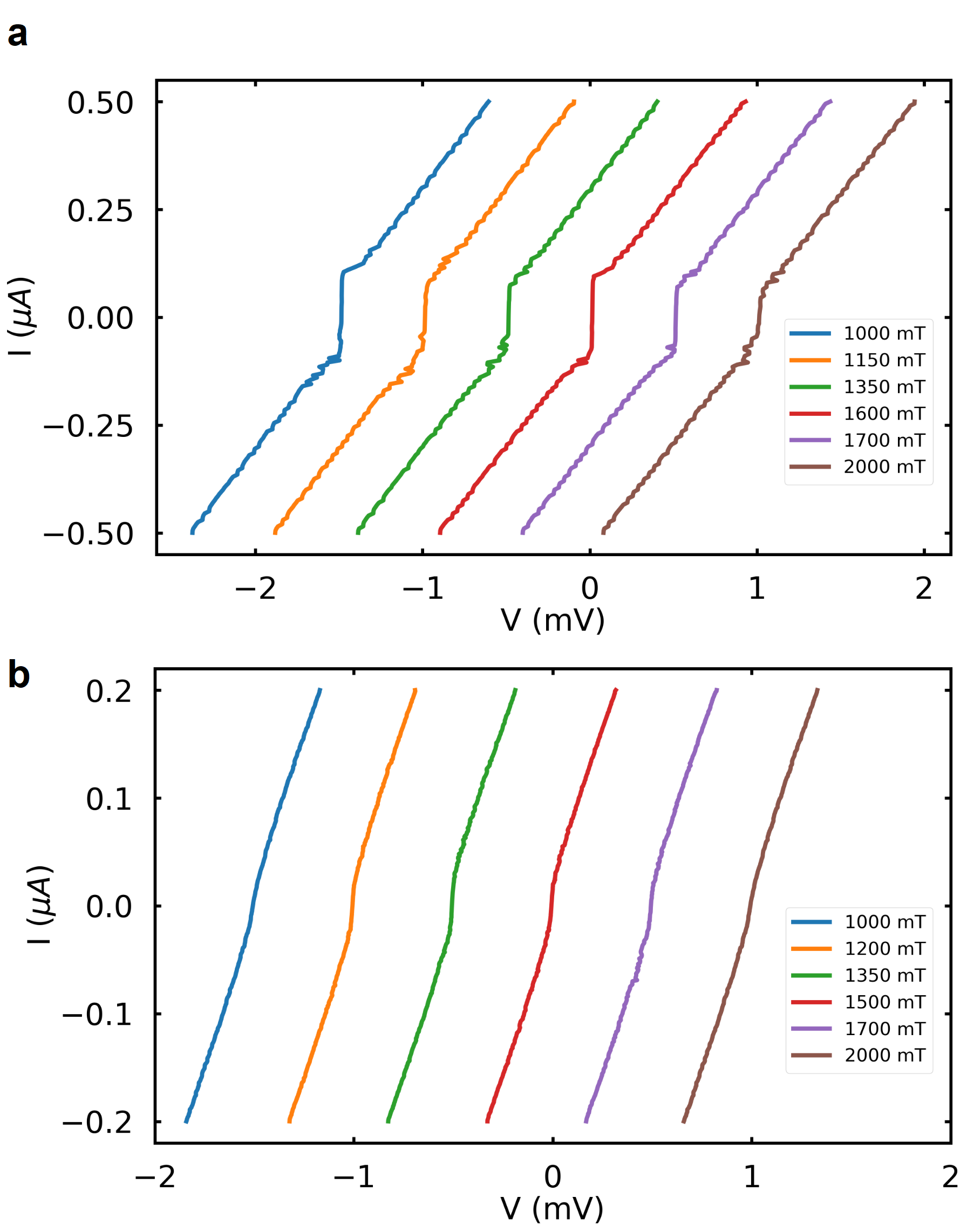}
	\caption{\label{fig:B_depend_gate}\textbf{a} Selected (I-V) curves at different $B$-field values $\geq$~1~T at $V_{\mathrm{sg}}=3.5$~V and \textbf{b} at $V_{\mathrm{sg}}=3.7$~V.}
\end{figure}

In the main text (see Fig. 2e), we compared the $B$-field dependence of the supercurrent of device B at finite temperature and finite gate voltage (with similar $I_{\mathrm{SW}}$ at $B$ =~0~T). We found that $B_{\mathrm{C}}$ decreases with increasing temperature, but not with increasing $V_{\mathrm{sg}}$. SFig.~\ref{fig:B_depend_gate}a,b shows some selected (I-V) curves at different $B$-field values $\geq$~1~T at $V_{\mathrm{sg}}=3.5$~V and $3.7$~V, respectively, separated in the x-axis for clarity.

\section{Temperature dependence of SCD}
\begin{figure}[ht!]
	\includegraphics[width=0.5\columnwidth]{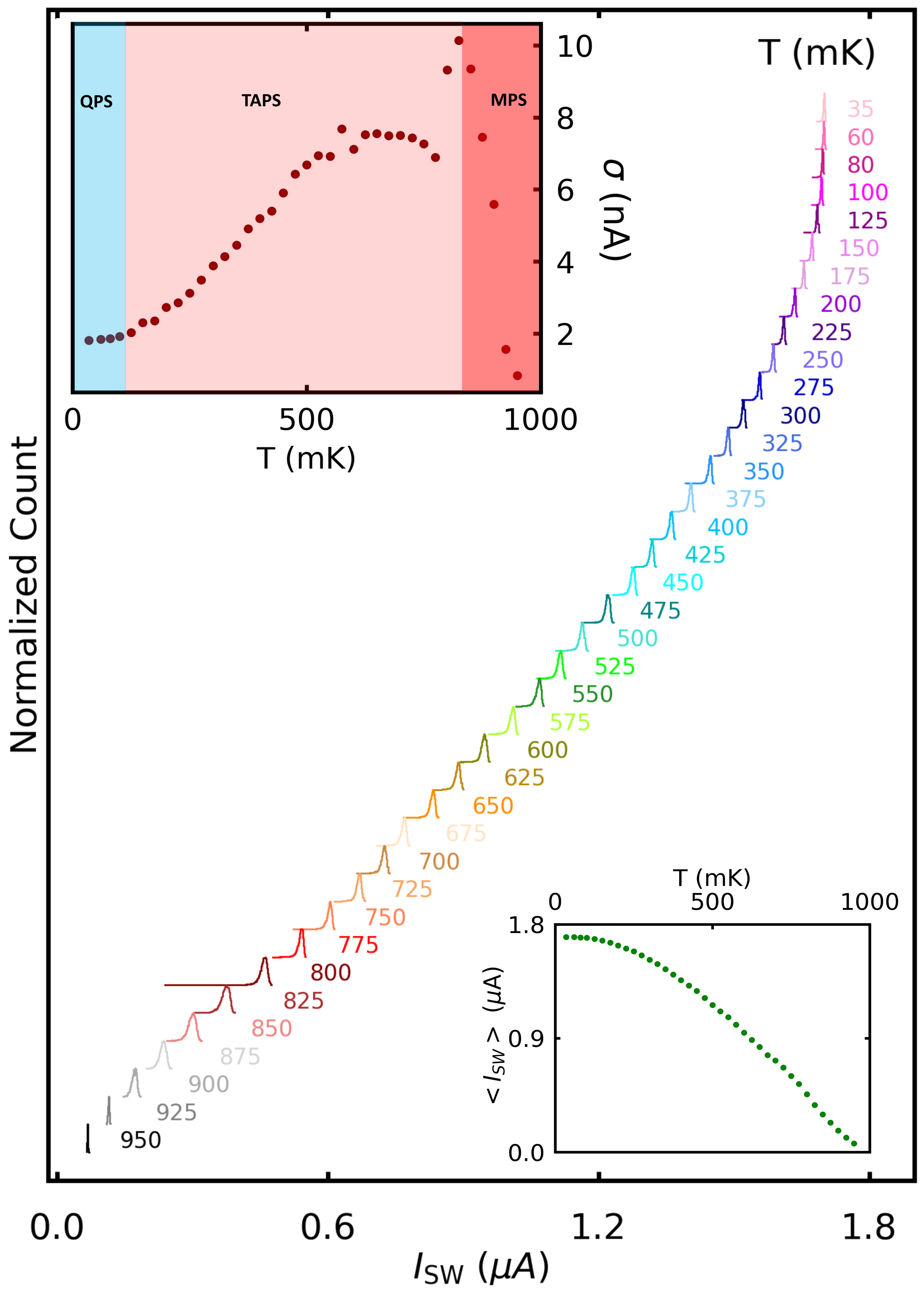}
	\caption{\label{fig:T_depend}\textbf{} Temperature dependence of SCDs for device C. In the top left, the standard deviation $\sigma$ of the SCDs is plotted as a function of temperature. The colored regions represent the quantum phase slips (QPS) regime (light blue), the thermally activated phase slips (TAPS) regime (pink), and the multiple phase slips (MPS) regime (orange-red). In the bottom right inset, the $\langle$$ I_{\mathrm{ SW }}$$\rangle$ of the SCDs is plotted as a function of temperature $T$.}
\end{figure}

The dependence of SCDs measured at elevated temperatures for device B is shown in SFig.~\ref{fig:T_depend}, in which SCDs measured at different temperatures at $B=0.1$~T and normalized to their maximum count are plotted and separated on the y-axis for clarity. The corresponding standard deviation $\sigma$ and mean value $\langle$$ I_{\mathrm{ SW }}$$\rangle$ are plotted as a function of temperature $T$ in the top left and bottom right insets, respectively.

The standard deviation $\sigma$ follows the expected conventional dependence as a function of $T$ (see Refs.\citenum{mccumber1970time,bezryadin2013superconductivity}), which can be divided into three distinct regions (different colored regions in the upper left inset) where the switching mechanism is attributed to quantum phase slips (QPS), thermally activated phase slips (TAPS), and multiple phase slips (MPS). Saturation of the width at the lowest temperatures can also originate from non- ideal thermalization of the electron bath, in which case this could still be part of the TAPS regime. Since this will not play an important role in our analysis, we will refer to this regime as the QPS regime for simplicity.

\section{dependence of SCD and escape rate on the current ramp speed}
\begin{figure}[ht!]
	\includegraphics[width=0.8\columnwidth]{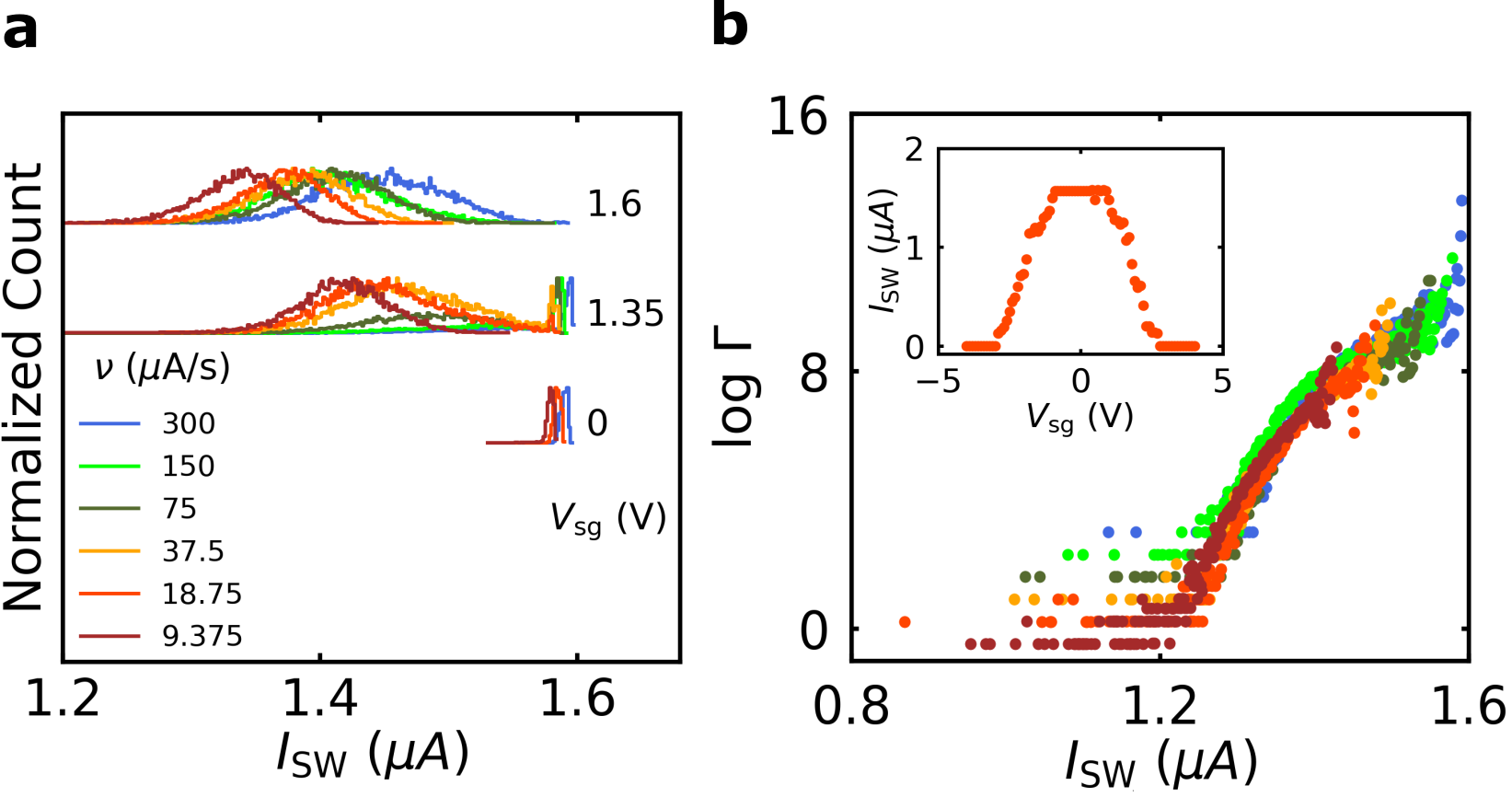}
	\caption{\label{fig:speed_depend}\textbf{a} For device C , the SCDs were measured at $V_{\mathrm{sg}}= $0$,$ $1.35,$ $1.6$ V, and for each value they were measured at different current ramp speeds $\nu_{\mathrm{I}}$. The SCDs are normalized by their maximum number of counts and separated on the y-axis. \textbf{b} Logarithm of escape rate $\Gamma$ as a function of $I_{\mathrm{ SW }}$ at different current ramp speeds $\nu_{\mathrm{I}}$ for SCDs measured at $V_{\mathrm{sg}} = 1.6$ V.}
\end{figure}

In this section, we discuss the dependence of SCDs on the current ramp speed $\nu_{\mathrm{I}}$ for device C in another cool-down in which the gate dependence of $I_{\mathrm{SW}}$ is changed ($V_{\mathrm{th}}$ and $V_{\mathrm{sg,C}}$) compared to that in the main text as shown in the inset of SFig.~\ref{fig:speed_depend}b. SFig.~\ref{fig:speed_depend}a shows the SCDs measured at three different values of $V_{\mathrm{sg}}$, and for each value they are measured at six different values of the current ramp speed $\nu_{\mathrm{I}}$. The values of $V_{\mathrm{sg}}$ are chosen so that the measured SCDs are at the beginning ($0$ V), within ($1.35$ V), and at the end ($1.6$ V) of the transition region between quantum tunneling (or thermal activation) and leakage-assisted phase slips.

At infinitely high values of $\nu$, the superconductor is expected to have no time to switch prematurely in the presence of external fluctuations, and it will switch at $I_{\mathrm{ SW }}$~=~$ I_{\mathrm{C}}$, at which the system prefers to set at the lower energy of the normal state \cite{bezryadin2013superconductivity}. Thus, when $\nu_{\mathrm{I}}$ increases from $9.375$ to $300$~$\mu$A/s, the $\langle$$I_{\mathrm{ SW }}$$\rangle$ of the histograms shifts to higher current values, as shown in SFig.~\ref{fig:speed_depend}a. Interestingly, the SCD e.g. at $1.35$~V appears as two overlapping probability distributions, where a complete switch between the two distributions depends on the value of $\nu_{\mathrm{I}}$. Despite the SCDs depend strongly on $\nu_{\mathrm{I}}$, the transformation of the measured probability distributions into the escape rate $\Gamma(I,T)$ is independent of $\nu_{\mathrm{I}}$ \cite{bezryadin2013superconductivity}.
The transformed $\Gamma(I,T)$ for $V_{\mathrm{sg}}=1.6$~V evaluated from  SCDs with different speeds  are plotted in SFig.~\ref{fig:speed_depend}b.

\section{Comparison between the influence of two opposite side gates for device A} 
\begin{figure}[H]
	\includegraphics[width=0.9\columnwidth]{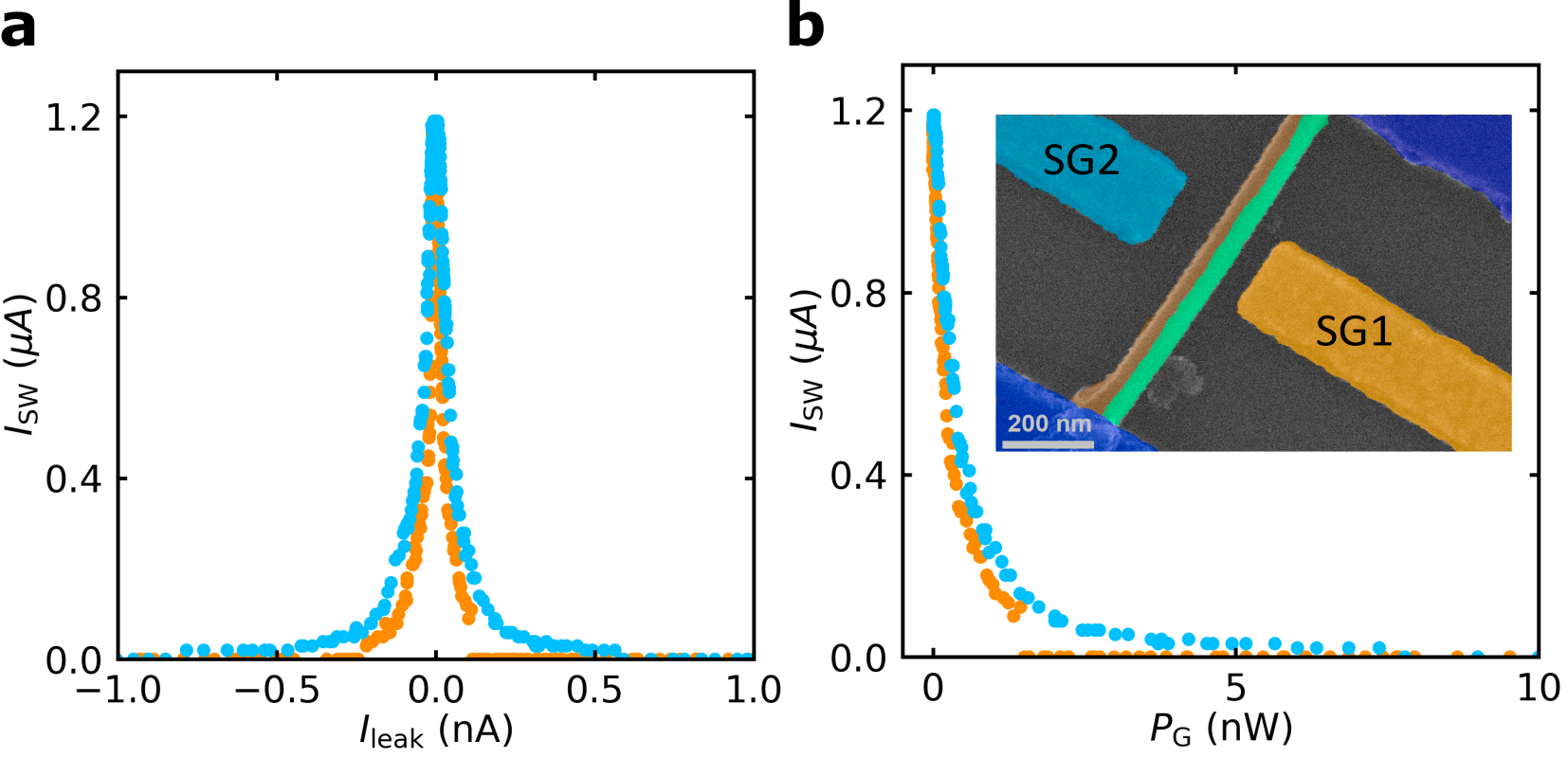}
	\caption{\label{fig:dual_gates}\textbf{a} Parametric curve between $I_{\mathrm{SW}}$ and $I_{\mathrm{leak}}$ for SG1 (orange curve) and for SG2 (light blue) of device A. \textbf{b} $I_{\mathrm{SW}}$ as a function of $P_{\mathrm{G}}$ for both gates. The inset show a false colored SEM image for the investigated device with two opposite side gates colored same as their corresponding curves.}
\end{figure}

Fig.1d and e in the main text show the dependence of $I_{\mathrm{SW}}$ on $V_{\mathrm{sg}}$ and the corresponding $I_{\mathrm{leak}}$ as a function of $V_{\mathrm{sg}}$ for two opposite side gates (SG1 and SG2) for device A. From these data, a parametric curve is plotted between $I_{\mathrm{SW}}$ and $I_{\mathrm{leak}}$ for the two gates in SFig.~\ref{fig:dual_gates}a. The plot shows that $I_{\mathrm{ SW }}$ is suppressed for both gates at the onset of $I_{\mathrm{leak}}$, despite it decreases at different $V_{\mathrm{th}}$. A larger $I_{\mathrm{leak}}$ is required to switch the device to the normal state at negative gate polarity than at the opposite polarity. Plotting $I_{\mathrm{SW}}$ as a function of $P_{\mathrm{G}}$ as shown in SFig.~\ref{fig:dual_gates}b, a better matching between the two curves is observed. We also noticed that for SG1, $P_{\mathrm{G,C}}$ is around 1.5 nW which is comparable to the switching power of the device $P_{\mathrm{n}}$~$\simeq$~ 1 nW.

\section{Investigation of device A in another cooldown}
\begin{figure}[H]
	\includegraphics[width=\columnwidth]{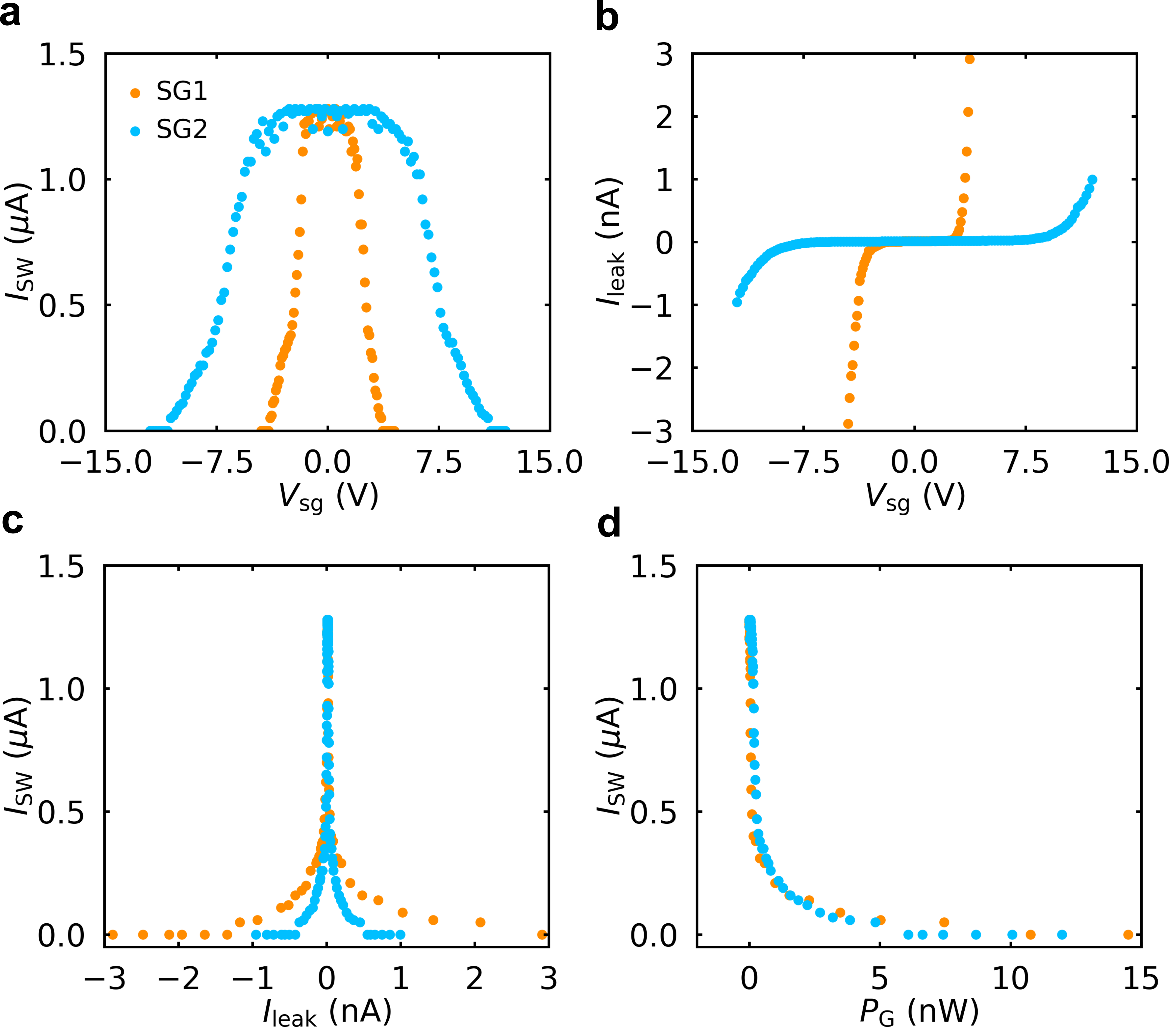}
	\caption{\label{fig:devA_2ndcool}\textbf{a} $I_{\mathrm{SW}}$ as a function of $V_{\mathrm{sg}}$ for SG1 (orange curve) and for SG2 (light blue curve) of device A in another cool-down.\textbf{b} $I_{\mathrm{leak}}$ as a function of $V_{\mathrm{sg}}$ for both gates. \textbf{c} Parametric curve between $I_{\mathrm{SW}}$ and $I_{\mathrm{leak}}$. \textbf{b} $I_{\mathrm{SW}}$ as a function of $P_{\mathrm{G}}$ for both gates.}
\end{figure}

By Investigating device A in another cool-down, the gate dependence of $I_{\mathrm{ SW }}$ (see SFig.~\ref{fig:devA_2ndcool}a) under the influence of SG1 (orange curve) and SG2 (light curve) gives an opposite situation to that obtained in the first cool-down (see Fig.1d,e in the main text). In this cool-down, SG1 suppresses $I_{\mathrm{ SW }}$ at lower $V_{\mathrm{th}}$ and $V_{\mathrm{sg,C}}$ than SG2. On the other hand, a corresponding increase in $I_{\mathrm{leak}}$ is observed at $V_{\mathrm{th}}$ for the two gates (see SFig.~\ref{fig:devA_2ndcool}b). 

As in the previous section, plotting a parametric curve between $I_{\mathrm{ SW }}$ and $I_{\mathrm{leak}}$ (see SFig.~\ref{fig:devA_2ndcool}c) for the two gates shows that $I_{\mathrm{ SW }}$ is suppressed at the onset of $I_{\mathrm{leak}}$. Moreover, a strong matching between the influence of the two gates is obtained when $I_{\mathrm{ SW }}$ is plotted as a function of $P_{\mathrm{G}}$, as shown in SFig.~\ref{fig:devA_2ndcool}d.

\section{dual-gate measurement}
\begin{figure}[H]
	\includegraphics[width=0.9\columnwidth]{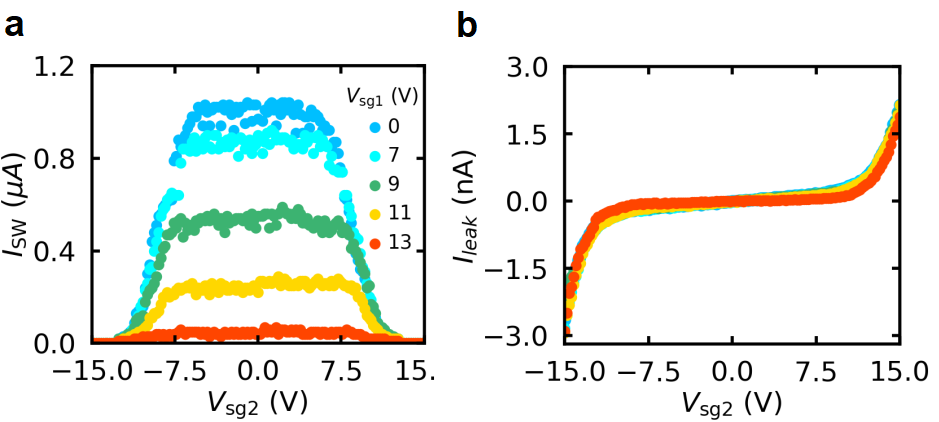}
	\caption{\label{fig:gates_comp}\textbf{a} $I_{\mathrm{SW}}$ and $I_{\mathrm{leak}}$ as a function of $V_{\mathrm{sg2}}$ at different values of $V_{\mathrm{sg1}}$ for device A \textbf{b} The corresponding $V_{\mathrm{sg}}$ versus $I_{\mathrm{leak}}$.}
\end{figure}

In this section, we will examine the effect of the two opposite side gates on the suppression of $I_{\mathrm{SW}}$ as well as on $V_{\mathrm{sg,C}}$. SFig.~\ref{fig:gates_comp}a shows $I_{\mathrm{SW}}$ as a function of $V_{\mathrm{sg2}}$ at different values of $V_{\mathrm{sg1}}$. For $V_{\mathrm{sg2}}=0$~V, $I_{\mathrm{SW}}$ decreases as expected with increasing $V_{\mathrm{sg1}}$, while with increasing $V_{\mathrm{sg2}}$, the dependence of $I_{\mathrm{ SW }}$ on $V_{\mathrm{sg2}}$ looks quite similar for all values of $V_{\mathrm{sg1}}$ and no change in $V_{\mathrm{sg2,C}}$ is observed. The independence $V_{\mathrm{sg,C}}$ for one of the gates from the influence of the other has already been reported in Ref.~\citenum{paolucci2019magnetotransport}, where the electric field was proposed as the origin of the gating effect. However, in our case a corresponding increase in the leakage current was observed as shown SFig.~\ref{fig:gates_comp}b.

\section{Measurements on different substrate }
\begin{figure}[H]
	\includegraphics[width=0.9\columnwidth]{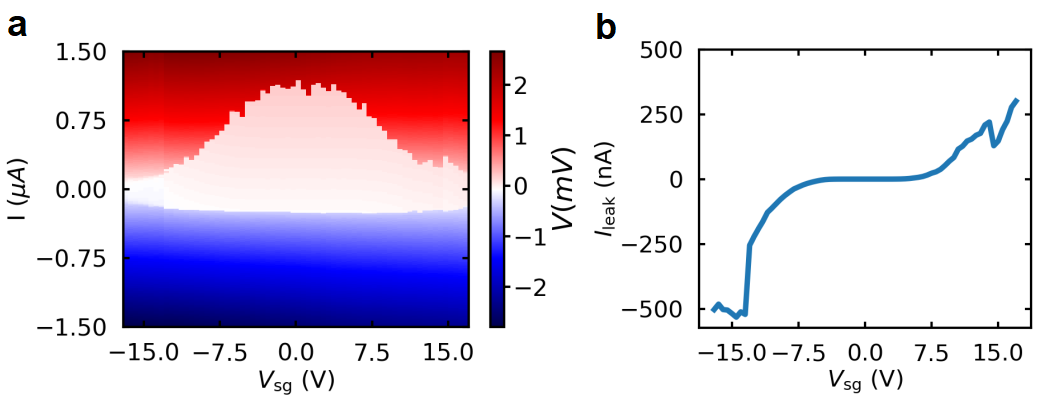}
	\caption{\label{fig:substrate_role}\textbf{a} $I-V$ curve as a function of $\pm$~$V_{\mathrm{sg}}$ of a Ta/InAs nanowire device fabricated on a sappire substrate. \textbf{b} The corresponding $I_{\mathrm{leak}}$ as a function of $\pm$~$V_{\mathrm{sg}}$. }
\end{figure}

    We have investigated the GCS in another Ta/InAs nanowire device with the same device configurations but fabricated on a sapphire substrate. Despite $I_{\mathrm{SW}}$ is not fully suppressed by $V_{\mathrm{sg}}$ in the $\pm$17 V window (see SFig.~\ref{fig:substrate_role}a), the corresponding $I_{\mathrm{leak}}$ (see SFig.~\ref{fig:substrate_role}b) and thus $P_{\mathrm{G}}$ is two orders of magnitude higher than for the devices fabricated on Si$^{\mathrm{++}}$/SiO$_{\mathrm{2}}$ substrate shown in the main text. A possible reason could be that in case of a  phonon mediated scenario the different type of the substrates allow different phonon generation. 

\bibliography{references}